\documentclass[runningheads]{llncs}
\usepackage[T1]{fontenc}
\usepackage{graphicx}
\usepackage{xcolor}
\usepackage[detect-all]{siunitx}
\usepackage{booktabs}
\usepackage{amsmath,amssymb,amsfonts}
\usepackage{multirow}

\usepackage{comment}

\usepackage[nolist]{acronym}

\usepackage{csquotes}

\usepackage{cleveref}
\crefname{figure}{Fig.}{Figs.}
\crefname{section}{Sec.}{Secs.}
\crefname{equation}{Eq.}{Eqs.}
\crefname{table}{Table}{Tables}

\begin{document}

\begin{acronym}
    \acro{rl}[RL]{Reinforcement Learning}
    \acro{ppo}[PPO]{Proximal Policy Optimization}
    \acro{drl}[DRL]{Deep Reinforcement Learning}
    \acro{morl}[MORL]{Multi-Objective Reinforcement Learning}
    \acro{mdp}[MDP]{Markov Decision Process}
    \acro{mpc}[MPC]{Model Predictive Control}
    \acro{1dof}[1-DoF]{one-degree-of-freedom}
    \acro{lqr}[LQR]{Linear-quadratic regulator}
    \acro{lqi}[LQI]{Linear-quadratic-integral regulator}
    \acro{lti}[LTI]{Linear Time-Invariant}
    \acro{mdp}[MDP]{Markov Decision Process}
    \acro{sb3}[SB3]{Stable Baselines3}
    \acro{trpo}[TRPO]{Trust Region Policy Optimization}
    \acro{icps}[ICPS]{Industrial \ac{cps}}
    \acro{cps}[CPS]{Cyber-Physical System}
    \acro{pid}[PID]{Proportional-integral-derivative}
    \acro{dof}[DoF]{Degree of Freedom}
    \acro{gp}[GP]{Gaussian Process}
    \acroplural{gp}[GPs]{Gaussian Processes}
    \acro{mobo}[MOBO]{Multi Objective Bayesian Optimization}
    \acro{uav}[UAV]{Unmanned Aerial Vehicle}
\end{acronym}

\title{Multi-Objective Reinforcement Learning for Energy-Efficient Industrial Control}
\titlerunning{MORL for Energy-Efficient Control}
%
\author{Georg Schäfer\inst{1,2,3} \and Raphael Seliger\inst{4} \and Jakob Rehrl\inst{1,2} \and Stefan Huber\inst{1,2} \and Simon Hirlaender\inst{3}}
\authorrunning{G. Schäfer et al.}
%
\institute{Josef Ressel Centre for Intelligent and Secure Industrial Automation \and
Salzburg University of Applied Sciences \and
Paris Lodron University of Salzburg \and
Kempten University of Applied Sciences
\\
\email{georg.schaefer@fh-salzburg.ac.at}}
\maketitle              
\begin{abstract}
Industrial automation increasingly demands energy-efficient control strategies to balance performance with environmental and cost constraints.
In this work, we present a multi-objective reinforcement learning (MORL) framework for energy-efficient control of the Quanser Aero~2 testbed in its one-degree-of-freedom configuration.
We design a composite reward function that simultaneously penalizes tracking error and electrical power consumption.

Preliminary experiments explore the influence of varying the energy penalty weight, $\alpha$, on the trade-off between pitch tracking and energy savings.
Our results reveal a marked performance shift for $\alpha$ values between \num{0.0} and \num{0.25}, with non-Pareto optimal solutions emerging at lower $\alpha$ values, on both the simulation and the real system.
We hypothesize that these effects may be attributed to artifacts introduced by the adaptive behavior of the Adam optimizer, which could bias the learning process and favor bang-bang control strategies.
Future work will focus on automating $\alpha$ selection through Gaussian Process-based Pareto front modeling and transitioning the approach from simulation to real-world deployment.

\end{abstract}
\section{Introduction}
Increasing energy prices and growing environmental awareness have become critical factors in industrial automation.
Therefore, scientists and practitioners are trying to find new solutions to reduce energy consumption in industrial automation while still achieving high performance.
This drives the need for advanced control strategies and hardware optimizations to effectively balance these competing goals~\cite{carabin_review_2017}.
This challenge is particularly evident in aerospace, where energy-efficient control strategies are crucial for battery-powered systems such as drones, ensuring extended flight durations~\cite{ramezani_energy-aware_2024}.

\ac{rl}, unlike classical methods such as \ac{mpc} and \ac{lqr}, does not require an explicit system model and can learn directly from interaction with the environment~\cite{schafer_comparison_2024}.
While our previous studies \cite{schafer_comparison_2024,ICPS} focused on optimizing trajectory tracking using \ac{rl}, it did not explicitly incorporate energy efficiency.
In this work, we extend this approach by using \ac{morl}, which enables the simultaneous optimization of multiple, often conflicting objectives \cite{hayes_practical_2022} like tracking accuracy and energy minimization through energy-aware reward shaping.

\subsection{Related Work}

\Ac{rl} has shown promise in handling nonlinear dynamics, uncertainty, and partial observability in various control tasks, including UAV stabilization~\cite{zhang_airpilot_2024} and trajectory tracking~\cite{jiang_quadrotor_2021,schafer_comparison_2024}.
However, most traditional \ac{rl} algorithms focus on a single objective~\cite{chunming_liu_multiobjective_2015}, while many industrial control problems require balancing multiple conflicting goals.

\Ac{morl} addresses this by either scalarizing multiple objectives using weighted sums~\cite{kim_adaptive_2006} or by directly handling vector-valued rewards via Pareto dominance to capture optimal trade-offs~\cite{vamplew_empirical_2011}.
In parallel, recent studies have integrated energy-awareness into \ac{rl} reward design through composite reward functions that balance performance metrics with energy consumption~\cite{ramezani_energy-aware_2024,jendoubi_multi-agent_2023,liao_energy_2022,yang_energy_2017}.
To our knowledge, however, no prior work has applied \ac{rl}-based Pareto modeling to simultaneously address energy minimization and tracking accuracy in twin-rotor or similar platforms.

\section{System Overview}
For the experimental evaluation, we employ the Quanser Aero~2 testbed in its 1-\ac{dof} configuration for energy-aware control of the system's pitch angle.
The system is operated using two motors with a single voltage input $u$.
The control signals are applied such that one motor receives $u_1 = u$ while the other receives $u_2 = -u$, with the input voltage bounded within \SIrange{-24}{24}{V}.
Experiments are conducted in a Simulink-based simulation interface with Python (for further details see~\cite{SMPS}).

\begin{figure}[ht]
  \centering \includegraphics[width=0.8\textwidth]{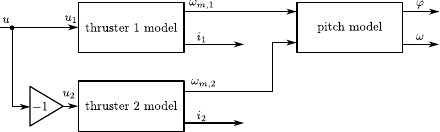}
  \caption{Block diagram of the system model.}
  \label{fig:blockd}
\end{figure}

The system model shown in \cref{fig:blockd} is used to predict the pitch angle $\varphi$ and motor currents $i$ of the Aero~2.
The thruster-blocks contain a DC motor model that neglects armature inductance and incorporates velocity-proportional load/friction torque, described by:

\begin{align}
  u&=i\,R+u_m,\quad u_m=k_m\,\omega_m\label{eq:armature}\\
  J_m\frac{d\omega_m}{dt}&=T_m-D_m\omega_m, \quad T_m=k_m\,i.\label{eq:thruster}
\end{align}

In \cref{eq:armature} and \cref{eq:thruster}, $\omega_m$ denotes the angular velocity of the thrusters, $R$ represents the armature resistance, $k_m$ is the motor constant, and $J_m$ is the inertia of the motor shaft and thrusters.
The back-EMF voltage and motor torque are denoted by $u_m$ and $T_m$, respectively, while $i$ represents the motor current.

The pitch model builds on the equations presented in~\cite{schafer_comparison_2024}, but departs from the assumption that thruster force is proportional to motor voltage.
Instead, thruster force is now modeled as proportional to the thrusters angular velocity.

\section{Problem Formulation}
The primary control objective is to track and maintain a desired pitch angle $r$ while minimizing energy consumption.
While previous work \cite{schafer_comparison_2024,ICPS} focused solely on tracking performance, this study extends the formulation by incorporating energy minimization.
To achieve this, we define a composite reward function within our \ac{rl} framework that balances tracking accuracy and energy efficiency:

\[
R_t = -(1 - \alpha) \cdot |\Delta_t| - \alpha \cdot P_t,
\]

where $\alpha \in [0, 1]$ is the energy penalty weight, $\Delta_t$ represents the normalized deviation from the target pitch, and $P_t$ denotes the normalized electrical power consumption.

Our methodology systematically analyzes the impact of different $\alpha$ values on the trade-off between pitch tracking and energy savings.
For the training procedure, five $\alpha$ values were initially evaluated with a step size of \num{0.25} (\num{0.0}, \num{0.25}, \num{0.5}, \num{0.75}, \num{1.0}).
Each training run was conducted for \num{500000} steps, with an evaluation phase every \num{10000} steps, during which the model was stored for further analysis.
Following the observation of a significant performance deviation when transitioning from $\alpha = 0.0$ to $\alpha = 0.25$, additional values (\num{0.05} and \num{0.10}) were tested.
Each experiment was executed five times with different random seeds, and performance was evaluated based on the mean and standard deviation of pitch deviation (in degrees) and power consumption (in watts).
For each value of $\alpha$, a single simulation-trained agent was deployed directly on the real system without any additional training or fine-tuning, allowing us to assess its performance in a real-world setting.

\section{Preliminary Results and Discussion}

\begin{figure}[ht]
    \centering
    \includegraphics[width=\linewidth]{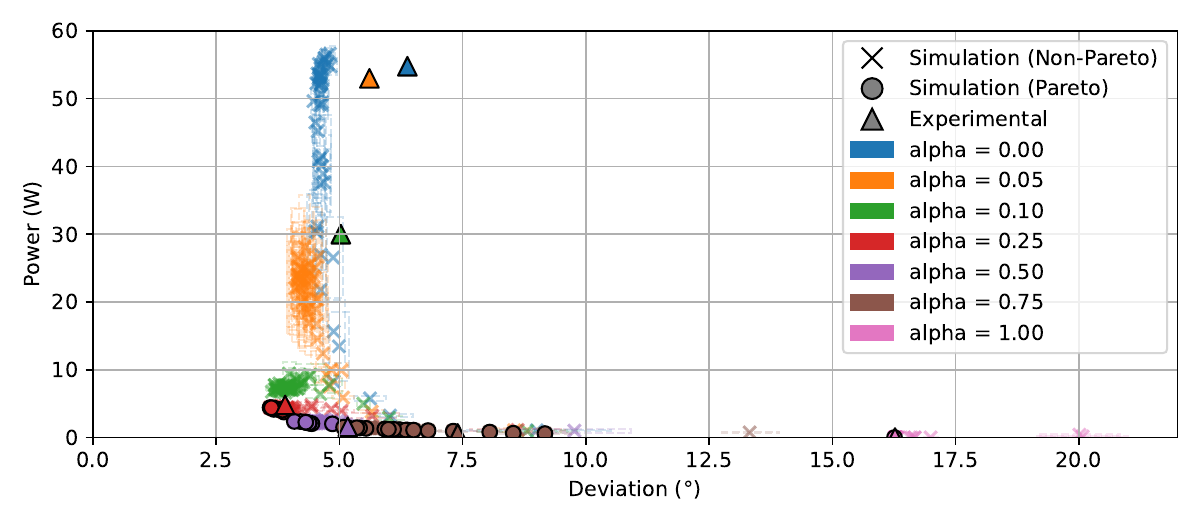}
    \caption{Pareto front of RL solutions for energy-efficient control.}
    \label{fig:pareto}
\end{figure}

The results, summarized in \cref{fig:pareto}, indicate a notable performance gap when adjusting the energy penalty weight.
Each data point represents the mean performance over five runs, with non-Pareto solutions marked by an \enquote{$\times$} and Pareto optimal points indicated by \enquote{$\circ$}.
Rectangles surrounding the markers denote the standard deviation in power consumption (watts) and tracking deviation (degrees) across the five trained agents, and colors correspond to the specific $\alpha$ values used.
Real-system evaluations are indicated by triangles.

The real-world performance mirrored the simulation trends: agents trained with $\alpha \leq 0.25$ exhibited aggressive control actions, causing large overshoots and persistent oscillations.
In contrast, for higher $\alpha$ values, the incorporation of the energy
penalty served as a regularizer, resulting in smoother control responses and
Pareto optimal performance. Prelmiminary experimens indicate that this effect is
less pronounced when Adam is replaced by SGD.
These observations suggest that the aggressive behavior at low $\alpha$ may be partially attributed to the adaptive nature of the Adam optimizer.
This highlights the need for further investigation into optimizer behavior, especially in multi-objective settings.

\section{Next Steps}
Our initial experiments have been conducted in simulation with manually chosen values for $\alpha$.
Future work aims to automate the selection of $\alpha$ values and further improve sample efficiency by leveraging \ac{gp} to model the Pareto front within the framework of \ac{mobo}.
This approach is expected to reduce the number of required training samples by efficiently approximating the trade-offs between tracking performance and energy consumption.
Additionally, future efforts will focus on transitioning from simulation-based training to real-world deployment on the Quanser Aero~2 hardware to validate the proposed methodology in a practical setting.

Moreover, the observed non-Pareto behavior for $\alpha \leq 0.25$ raises intriguing questions regarding the utilization of adaptive optimizers, like Adam, for multi-objective goals.
Investigating and mitigating these optimizer-induced artifacts will be a crucial aspect of our future research.

\begin{credits}
\subsubsection{\ackname} Financial support for this study was provided by the Christian Doppler Association (JRC ISIA), the corresponding WISS Co-project of Land Salzburg, the European Interreg Österreich-Bayern project BA0100172 AI4GREEN.

\end{credits}
%
%
%
%
\bibliographystyle{IEEEtran}
\bibliography{references}
\end{document}